\documentclass[aps,prl,showpacs,twocolumn,groupedaddress]{revtex4}

\usepackage{graphicx}% Include figure files
\usepackage{dcolumn}% Align table columns on decimal point
\usepackage{bm}% bold math

\begin{document}

\preprint{APS/123-QED}

\title{Spin-Liquid State in the \textit{S} = 1/2 Hyperkagome Antiferromagnet Na$_4$Ir$_3$O$_8$}% Force line breaks with \\

%\author{Ann  Author}
% \altaffiliation[Also at ]{Physics Department, XYZ University.}%Lines break aut%omatically or can be forced with \\
%\affiliation{%
%\affiliation{

\author{Yoshihiko Okamoto$^{1,*}$, Minoru Nohara$^{2}$, 
%Hiroyuki Mitamura$^4$, 
Hiroko Aruga-Katori$^{1}$, and Hidenori Takagi$^{1,2}$}
\affiliation{
$^{1}$RIKEN (The Institute of Physical and Chemical Research), 2-1 Hirosawa, Wako, Saitama 351-0198, Japan\\
%$^{2}$CREST, Japan Science and Technology Agency (JST), Japan\\
$^{2}$Department of Advanced Materials, University of Tokyo and CREST-JST, 5-1-5 Kashiwanoha, Kashiwa, Chiba 277-8561, Japan\\
%$^{4}$Institute for Solid State Physics, University of Tokyo, 5-1-5 Kashiwanoha, Kashiwa, Chiba 277-8581, Japan
}

\date{\today}% It is always \today, today,

\begin{abstract}
A spinel related oxide, Na$_4$Ir$_3$O$_8$, was found to have a three dimensional
network of corner shared Ir$^{4+}$ (\textit{t}$_{2g}$$^5$) triangles.
This gives rise to an antiferromagnetically coupled $S$ = 1/2 spin system
formed on a geometrically frustrated hyperkagome lattice.
Magnetization $M$ and magnetic specific 
heat $C_\textrm{m}$ data showed the absence of long range 
magnetic ordering at least down to 2 K. 
The large $C_\textrm{m}$ at low temperatures
%, which shows a power law decay with temperature and 
is independent of applied magnetic field up to 12 T, 
in striking parallel to
the behavior seen in triangular and kagome antiferromagnets reported to have a
spin-liquid ground state.
These results strongly suggest that the ground state of 
Na$_4$Ir$_3$O$_8$ is a three dimensional manifestation of a spin liquid.
\end{abstract}

\pacs{Valid PACS appear here}% PACS, the Physics and Astronomy
\maketitle

The experimental realization of a quantum spin liquid in geometrically
frustrated magnets has been 
one of the biggest challenges in the field of magnetism 
since Anderson proposed resonating valence bond theory~\cite{RVB}
for antiferromagnetically coupled $S$ = 1/2 spins on a triangular lattice.
Geometrical frustration in magnets
arises from the incompatibility of local spin-spin interactions, 
which gives rise to macroscopic degeneracy of the ground state.
Possible playgrounds for this include
triangular, kagome, pyrochlore and garnet lattices essentially consisting
of networks of triangles.
%The frustrations are known to be the strongest for $S$ = 1/2 Heisenberg spins.
In real materials, however, it is not easy to prevent spin ordering at
substantially lower temperatures than the Curie-Weiss temperature $\theta_{\mathrm{W}}$.
%, the mean field transition temperature.
This is because the spin degeneracy can be lifted by coupling with
the other degrees of freedom such as the orbitals, lattice and charges. 
Such an interplay between the frustrated spins, orbitals and lattice,
for example, can be realized in the trimer singlet formation in the $S$ = 1
triangular LiVO$_2$~\cite{LiV,LiV2} with orbital ordering or 
the spin-Jahn-Teller transition in the $S$ = 3/2 
pyrochlore ZnCr$_2$O$_4$~\cite{ZnCr}.
In addition, only a minute amount of disorder can strongly influence
the spin-liquid state in geometrically frustrated magnets
and may give rise to the formation 
of a glassy state of spins.

The most likely candidate for the realization of a spin-liquid ground state
has been the two dimensional
kagome antiferromagnet SrCr$_{9p}$Ga$_{12-9p}$O$_{19}$
($S =$ 3/2)~\cite{SCG,SCGO}.
It does not show any evidence for long range ordering down to
% the lowest temperature $\sim$ 
100 mK,
and a large and field independent magnetic specific heat was observed which
was ascribed to spin-liquid contributions.
Nevertheless, the strong spin glass-like behavior at low temperatures
%, very likely due to site disorder, 
instills a certain ambiguity in identifying the spin-liquid state.
Recently, a new generation of spin-liquid compounds has emerged, the $S$ = 1/2 triangular magnet $\kappa$-(ET)$_2$Cu$_2$(CN)$_3$~\cite{et}, an organic Mott insulator, and the $S$ = 1 triangular magnet NiGa$_2$S$_4$~\cite{NiGa}. 
They were reported to have a spin-liquid ground state or at least a robust liquid phase down to 100 mK.
Their magnetic and thermal properties are %reported to have a
in striking parallel to those of SrCr$_{9p}$Ga$_{12-9p}$O$_{19}$ but the 
disorder effect appears to be much weaker.

Here we report on a three dimensional analogue of these
two dimensional spin liquids. Na$_4$Ir$_3$O$_8$ was first reported as an
unidentified phase in the Na-Ir-O ternary system by McDaniel~\cite{NaIr}.
We find that it is isostructural to Na$_4$Sn$_3$O$_8$~\cite{NaSn} and
that a $S$ = 1/2 hyperkagome system, consisting of low spin \textit{d}$^5$ Ir$^{4+}$ ions, is realized in Na$_4$Ir$_3$O$_8$. The magnetization and specific heat measurements on the ceramic samples indicate that $S$ = 1/2 spins are highly frustrated and remain in a liquid state down to the lowest temperature measured.

Polycrystalline samples of Na$_4$Ir$_3$O$_8$ were prepared by a solid-state reaction.
Stoichiometric amounts of Na$_2$CO$_3$ and IrO$_2$ were mixed, and the mixture was calcined at 750$^\circ$C for 18 h. We added 5 \% excess of Na$_2$CO$_3$ to compensate the loss of Na during the calcination. The product was finely ground, pressed into a pellet, sintered at 1020$^\circ$C for 22 h on gold foil, and then quenched in air. 
Powder x-ray diffraction (XRD) data showed that the powders were single
phase.
The crystal structure was determined by performing Rietveld analysis on the powder XRD data using RIETAN-2000 program~\cite{Rietan}. 
Thermodynamic and magnetic properties were measured by a
Physical Properties Measurement System (Quantum Design) and
a Magnetic Properties Measurement System (Quantum Design). 

\begin{figure}
\includegraphics[width=8cm]{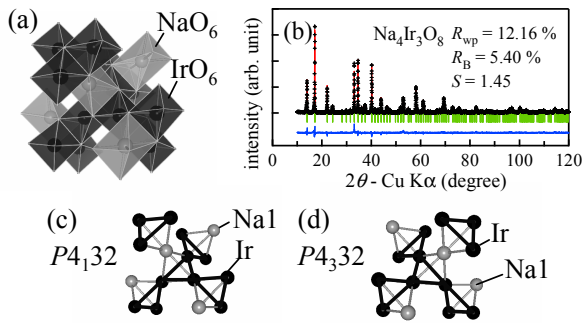}% Here is how to import EPS art
\caption{\label{fig1}(a) Crystal structure of Na$_4$Ir$_3$O$_8$ with
the space group $P4_{1}32$.
Among the three Na sites, only Na1 site is shown for clarity.
Black and gray octahedra represent IrO$_6$ and NaO$_6$ respectively.
The spheres inside the octahedra represent Ir and Na atoms
and oxygens occupy all the corners.
(b) The x-ray diffraction pattern of Na$_4$Ir$_3$O$_8$ at
room temperature. The crosses indicate the raw data and the solid line
indicates the spectrum calculated based on the refinement
using $P4_{1}32$.
(c) and (d) Hyperkagome Ir and Na sublattice derived from the structure of
Na$_4$Ir$_3$O$_8$ with the space group $P4_{1}32$ (c)
and $P4_{3}32$ (d). These two structures with different chirality
are indistinguishable by conventional x-ray diffraction, giving the identical
 result in refinement.
%, shown in (a).
%(d) Ir and Na sublattice derived from the structure of Na$_4$Ir$_3$O$_8$
%with the space group $P4_{3}32$.
}
\end{figure}
\begin{table}
\caption{\label{table1}Atomic parameters obtained by refining x-ray powder diffraction for Na$_4$Ir$_3$O$_8$ at room temperature with a space group $P4_{1}32$.
The cubic lattice constant is $a$ = 8.985 \AA.
$g$ of Na2 and Na3 are fixed to 0.75 according to Ref.~\cite{NaSn}.}
\begin{ruledtabular}
\begin{tabular}{ccccccc}
	& & $x$ & $y$ & $z$ & $g$ & $B$ (\AA) \\ \hline
	Ir & 12\textit{d} & 0.61456(7) & $x$ + 1/4 & 5/8 & 1.00 & 0.15 \\		
	Na1 & 4\textit{b} & 7/8 & 7/8 & 7/8 & 1.00 & 2.6 \\ 
	Na2 & 4\textit{a} & 3/8 & 3/8 & 3/8 & 0.75 & 2.6 \\
	Na3 & 12\textit{d} & 0.3581(8) & $x$ + 1/4 & 5/8 & 0.75 & 2.6 \\
	O1 & 8\textit{c} & 0.118(11) & $x$ & $x$ & 1.00 & 0.6 \\
	O2 & 24\textit{e} & 0.1348(9) & 0.8988(8) & 0.908(11) & 1.00 & 0.6 \\ 
\end{tabular}
\end{ruledtabular}
\end{table}

We were able to refine the powder XRD pattern with the
cubic Na$_4$Sn$_3$O$_8$ structure ($P4_{1}32$ or $P4_{3}32$)~\cite{NaSn}.
The result of this refinement is summarized in Table I and Fig.~1 (b).
The structure of Na$_4$Ir$_3$O$_8$, shown in Fig.~1 (a), is derived from
those of spinel oxides (\textit{AB}$_2$O$_4$), which can be intuitively
demonstrated by rewriting the chemical formulae as
(Na$_{1.5}$)$_1${(Ir$_{3/4},$ Na$_{1/4}$)$_2$O$_4$. 
The \textit{B}-sublattice of spinel oxides forms the so-called pyrochlore lattice,
a network of corner shared tetrahedra. 
In Na$_4$Ir$_3$O$_8$, each tetrahedron in the \textit{B}-sublattice is occupied
by three Ir and one Na (Na1).
These Ir and Na atoms form an intriguing ordering pattern as shown in Fig.~1 (c), giving rise to a network of corner shared Ir triangles, called 
a hyperkagome lattice~\cite{Ramirez}.
All the Ir sites and Ir-Ir bonds are equivalent and, therefore,
strong geometrical frustration is anticipated. 
Hyperkagome is also realized in the \textit{A}-sublattice of the
garnet \textit{A}$_3$\textit{B}$_5$O$_{12}$ but these it is distorted.
It might be interesting to infer here that there exists a chirality
in this hyperkagome lattice and that the two
structures $P4_{1}32$ [Fig.~1 (c)] and $P4_{3}32$ [Fig.~1 (d)]
have different degenerate chiralities.
Na$_{1.5}$ in Na$_{1.5}$(Ir$_{3/4},$ Na$_{1/4}$)$_2$O$_4$ occupies the
octahedral \textit{A} site rather than the tetrahedral \textit{A} site normally
occupied in a conventional spinel structure~\cite{NaSn}.
We refined the structure by assuming two Na positions, Na2 and Na3, 
in the octahedral \textit{A}-site with 75 \% occupation following Ref.~\cite{NaSn}.
%octahedra with 75 \% occupation following Ref.~\cite{NaSn}.
%There remains a certain ambiguity in the refinement of the Na2 and Na3 sites
%because of the small scattering factor compared with Ir.

\begin{figure}
\includegraphics[width=7cm]{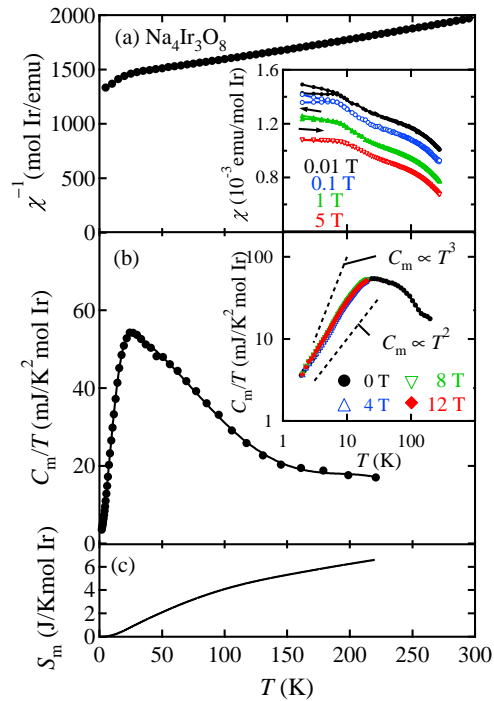}% Here is how to import EPS art
\caption{\label{fig2} 
Temperature dependence of the inverse
magnetic susceptibility $\chi^{-1}$ under 1 T (a),
magnetic specific heat $C_{\mathrm{m}}$ divided by
temperature $T$ (b) and magnetic entropy $S_{\mathrm{m}}$ (c) of
polycrystalline Na$_4$Ir$_3$O$_8$.
To estimate $C_{\mathrm{m}}$, data for Na$_4$Sn$_3$O$_8$ is used as
a reference of the lattice contribution.
Inset: (a) Temperature dependence of magnetic susceptibility $\chi$ of
Na$_4$Ir$_3$O$_8$ in various fields up to 5 T. 
For clarity, the curves are shifted
by 3, 2 and 1 $\times$ $10^{-4}$ emu/mol Ir for 0.01, 0.1 and
1 T data respectively. 
(b) $C_{\mathrm{m}}$/$T$ vs $T$ of Na$_4$Ir$_3$O$_8$ in various fields
up to 12 T. Broken lines indicate $C_{\mathrm{m}}$ proportional
to $T^{2}$ and $T^{3}$ respectively.
}
\end{figure}

Ir in this compound is tetravalent with five electrons
in 5\textit{d} orbitals.
Because of the octahedral coordination with the oxygens
and the large crystal field splitting effect expected
for 5\textit{d} orbitals,
it is natural for Ir$^{4+}$ to have a low spin
(\textit{t}$_{2g}$$^5$) state with $S$ = 1/2.
The electrical resistivity $\rho$ of a ceramic sample at room temperature
was $\sim$10 $\Omega$cm, followed by a thermally activated increase
with an activation energy of 500 K with decreasing temperature.
This, together with the magnetic properties described 
below, indicates
that Na$_4$Ir$_3$O$_8$ is a $S$ = 1/2 Mott insulator formed on a hyperkagome lattice.

The temperature dependent magnetic susceptibility $\chi$($T$),
shown in Fig.~2 (a), indicates that Na$_4$Ir$_3$O$_8$ is indeed a frustrated
$S$ = 1/2 system with a strong antiferromagnetic interaction.
In the $\chi^{-1}$ vs $T$ plot in Fig.~2 (a),
Curie-Weiss like behavior can be seen. 
The Curie-Weiss fit around room temperature yields a large
antiferromagnetic Curie-Weiss constant $\theta$$_{\mathrm{W}}$ $\sim $ 650 K
and an effective moment $p_{\mathrm{eff}}$ = 1.96 $\mu$$_{\mathrm{B}}$,
which is slightly larger than those expected for $S$ = 1/2 spins. 
In geometrically frustrated antiferromagnets, it is known that the Curie-Weiss
behavior expected above $T$ = $\theta$$_{\mathrm{W}}$ persists even
below $\theta$$_{\mathrm{W}}$.
The observed Curie-Weiss behavior of $\chi$($T$)
below $\theta$$_{\mathrm{W}}$ is consistent with the presence of
$S$ = 1/2 antiferromagnetic spins on a frustrated hyperkagome lattice.
The large antiferromagnetic interaction inferred 
from $\theta$$_{\mathrm{W}}$ is supported by the observation of
a magnetization linear with magnetic field at least up to 40 T without any sign
of saturation at 4.2 K~\cite{HF}.

The geometrical frustration in the $S$ = 1/2
hyperkagome antiferromagnet is extremely strong and,
indeed, we do not find any anomaly indicative of long range ordering
in the susceptibility at least down to 2 K, which is two orders of
magnitude lower than $\theta$$_{\mathrm{W}}$ $\sim$ 650 K.
We also note that a neutron diffraction measurement at 10 K did not
detect any signature of ordering~\cite{neutron}.
These strongly suggest that a spin-liquid state is indeed realized in this three
dimensional $S$ = 1/2 frustrated magnet.
As shown in the inset of Fig.~2 (a), a trace of spin glass like 
contribution with $T_{\textrm{g}}$ = 6 K is observed.
The difference between zero-field cooling and field cooling
magnetization, however,
is less than 10 \% of the total magnetization. This
hysteresis does not represent a contribution from the majority of spins.
%(see also Fig.~4 (a)).}
%clearly the 
%glassy contributions in the susceptibility, 
%with hysteresis between zero-field cooling and field cooling,
%are at most 10 \% of the total susceptibility and should be ascribed
%to an extrinsic origin (see also Fig.~4 (a)).
%This is also supported by the nonmagnetic impurity effect described later.
The glassy component
% in the magnetization 
becomes negligibly small at high fields above 1 T, 
relative to the other contributions.
In the high field susceptibility data that most likely represents the bulk, 
we see the susceptibility tend to saturate and approach a finite value as $T \to 0$. 
This strongly suggests that the majority of the system remains a paramagnetic spin liquid at
least down to 2 K. 

\begin{figure}
\includegraphics[width=6cm]{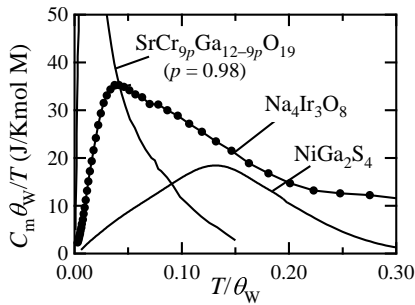}% Here is how to import EPS art
\caption{\label{fig3} Comparison of the normalized magnetic specific heat
of Na$_4$Ir$_3$O$_8$ with those of other frustrated antiferromagnets
SrCr$_{9p}$Ga$_{12-9p}$O$_{19}$ ($p$ = 0.98)~\cite{SCGO} and NiGa$_2$S$_4$~\cite{NiGa}.
M in the unit of vertical axis denotes magnetic element Ir, Cr and Ni 
for Na$_4$Ir$_3$O$_8$, SrCr$_{9p}$Ga$_{12-9p}$O$_{19}$ and
NiGa$_2$S$_4$ respectively.
Temperature $T$ is normalized by the Curie-Weiss
constant $\theta_{\mathrm{W}}$ for comparison.
}
\end{figure}
The specific heat data provides further evidence for a spin-liquid state.
The magnetic specific heat was estimated by subtracting 
the specific
heat of nonmagnetic Na$_4$Sn$_3$O$_8$ as a lattice contribution. 
Because of the subtraction, the data at high temperatures
above $\sim$100 K, where the lattice contribution dominates
the specific heat, are subject to certain ambiguity. 
The $T$-dependent magnetic specific heat $C_{\mathrm{m}}$ of Na$_4$Ir$_3$O$_8$ is
plotted as $C_{\mathrm{m}}$/$T$ in Fig.~2 (b). 
We observe only a broad peak with its maximum around $\sim$30 K
and any anomaly indicative of long range ordering is absent. 
The magnetic entropy, estimated by integrating $C_{\mathrm{m}}$/$T$-$T$ data shown in Fig.~2 (c), is as large as $\sim$4.5 J/molK per Ir at 100 K ($\ll \theta_{\mathrm{W}}$ = 650 K), which is 70-80 \% of the total spin entropy $R$ln 2 = 5.7 J/molK. 
The quenching of spin entropy at lower temperature than the Weiss temperature
$\theta_{\mathrm{W}}$ is a hallmark of frustrated systems, often refered to as a 
downshift of entropy.
Comparing with other frustrated systems in Fig.~3,
the downshift with respect to
 the Curie-Weiss temperature is much more
significant than in the two dimensional $S$ = 1 
NiGa$_2$S$_4$~\cite{NiGa} but less significant than in the
two dimensional $S$ = 3/2 kagome SrCr$_{9p}$Ga$_{12-9p}$O$_{19}$~\cite{SCGO}. 

\begin{figure}
\includegraphics[width=6cm]{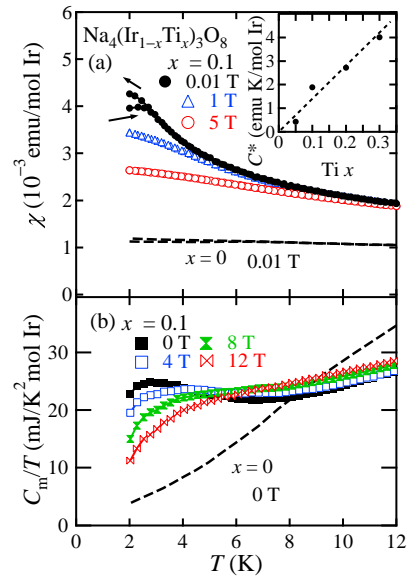}% Here is how to import EPS art
\caption{\label{fig4} (a) Temperature dependence of the magnetic susceptibility
$\chi$($T$) of polycrystalline Na$_4$(Ir$_{1-x}$Ti$_x$)$_3$O$_8$ ($x$ = 0.1)
in magnetic fields up to 5 T, compared to the data of $x$ = 0 under 0.01 T.
(b) Magnetic specific heat $C_{\mathrm{m}}$ of the $x$ = 0.1 sample 
plotted as $C_{\mathrm{m}}$/$T$ vs $T$. 
The broken line indicates $C_{\mathrm{m}}$/$T$ of $x$ = 0 under zero field. 
inset: Orphan spin Curie constant $C^*$ of Na$_4$(Ir$_{1-x}$Ti$_x$)$_3$O$_8$ 
(0 $\le x \le$ 0.3), defined by Schiffer and Daruka~\cite{orphan}. 
}
\end{figure}

As seen in the inset of Fig.~2 (b), the magnetic specific heat was found to be
surprisingly independent of applied magnetic fields up to $H$ = 12 T,
which corresponds to $\mu_{\mathrm{B}}H$/$k_{\mathrm{B}}$ $\sim$ 8 K. 
This suggests that the low energy spin excitation, seen as a large
magnetic specific heat at low temperature, has nothing to do with the
glassy contribution with the characteristic energy scale
of $T_{\mathrm{g}} \sim$ 6 K but derives from frustrated spins
strongly coupled antiferromagnetically. 
This field independence is universally observed in geometrically frustrated
magnets proposed to have a spin-liquid ground state~\cite{SCGO,NiGa}, 
providing a further support for a similar state in Na$_4$Ir$_3$O$_8$.

We also found that nonmagnetic 
Ti$^{4+}$ can be substituted partially for Ir$^{4+}$.
As shown in Fig.~4, the introduction of ``non-magnetic'' Ti$^{4+}$ impurities
gives rise to a localized magnetic moment, 
which manifests itself as a Curie-like contribution in the susceptibility, 
roughly scaled by the number of Ti$^{4+}$ ($S$ = 1/2 per 3Ti$^{4+}$). 
This is induced by the so-called orphan spin, 
and is again analogous to the other spin-liquid systems~\cite{orphan}. 
These localized magnetic moments simultaneously give rise to a drastic
shift of the magnetic specific heat to even lower temperatures
as shown in Fig.~4 (b).
This low-temperature specific heat in Ti$^{4+}$ doped
samples, however, is strongly magnetic field dependent [Fig.~4 (b)],
indicating that it has a physically distinct origin from those of
the nominally pure compound.
Incidentally, the Curie-like contribution induced by Ti$^{4+}$ is
accompanied by 
an enhanced hysteresis at low temperatures [Fig.~4 (a)],
which may support the idea that the glassy contribution seen in the nominally
pure compound originates from a small amount of impurity or disorder.

These experimental results all point to a spin-liquid ground state in
Na$_4$Ir$_3$O$_8$. 
Recent theoretical calculations using the large $N$ mean field theory 
indeed support spin-liquid formation on a hyperkagome lattice~\cite{YBK}.
However, there remain many issues and puzzles on the novel spin-liquid state of Na$_4$Ir$_3$O$_8$ which should be tackled urgently.
Firstly, the orbital state of Ir$^{4+}$ should be clarified in understanding 
the spin-liquid state of Na$_4$Ir$_3$O$_8$, because orbital ordering often 
results in anisotropic spin coupling and hence suppresses frustration. 
Taking a close look at the atomic coordination in Table I,
one notices that, because of chemical pressure from the large Na$^{+}$ ion in Ir$_3$Na tetrahedron,
the IrO$_6$ octahedra are distorted and elongate 
towards the center of the Ir$_3$Na tetrahedra.
This de-stabilizes the $a_{1\textrm{g}}$ orbital ($\{|xy\rangle + |yz\rangle + |zx\rangle\}/\sqrt{3}$) pointing towards the center of the Ir$_3$Na tetrahedra. 
We may speculate that the $S$ = 1/2 on Ir$^{4+}$ has primarily $a_{1\textrm{g}}$ character. If this is the case, the interactions between the $S$ = 1/2 spins originate from exchange coupling through the overlap of $a_{1\textrm{g}}$ orbitals. All the nearest neighbor interactions then should be equivalent and Heisenberg-like, consistent with the presence of strong geometrical frustration.

Secondly, the effect of spin-orbit coupling should be considered. 
Since Ir is a 5\textit{d} element, the spin-orbit coupling is
likely to be much larger than
in 3\textit{d} and 4\textit{d} elements. 
The large spin-orbit coupling will give rise to a spin anisotropy
and can reduce the frustration to a certain extent. 
It is likely from the experimental observation here, however, 
that this effect is not sufficient to suppress 
the spin-liquid state completely.

Finally, the origin of the unusual temperature dependence of the magnetic
specific heat $C_\textrm{m}$ is worthy of further exploration.
$C_\textrm{m}$ at low temperatures shows
a weaker temperature dependence
than $T^3$ at least down to 2 K [see the inset of Fig.~2 (b)]. 
This approximately $T^2$-behavior is in striking parallel with 
the behavior found in the $S$ = 1 triangular 
NiGa$_2$S$_4$~\cite{NiGa} and 
the $S$ = 3/2 kagome SrCr$_{9p}$Ga$_{12-9p}$O$_{19}$~\cite{SCGO}. 
In those two dimensional frustrated magnets, the $T^2$-dependence
of $C_\textrm{m}$($T$) at low temperatures may be interpreted
as the presence of a 2D magnon-like dispersion~\cite{SCGO,NiGa}. 
The hyperkagome lattice, however, is a three dimensional system
and it is not obvious at all why low temperature specific heat shows such a peculiar temperature dependence.

In conclusion, we have demonstrated that a spinel related oxide
Na$_4$Ir$_3$O$_8$ has an intriguing Ir-sublattice,
due to ordering of Na and Ir in the spinel B-site and that a $S$ = 1/2
hyperkagome antiferromagnet is realized in this oxide. 
The magnetization and 
specific heat data collectively suggest that the ground state is a spin
liquid state due to strong geometrical frustration. 
This is the first demonstration of a $S$ = 1/2 spin-liquid ground state 
in a three dimensional magnet and, we believe, provides a new, and fascinating playground for quantum magnetism.

We thank D.\ I.\ Khomskii, N.\ E.\ Hussey, T.\ Arima, S.\ Onoda, S.\ Shamoto and
H.\ Mitamura for stimulating discussion.
This work was partly supported by a Giant-in-Aid for Scientific Research, from the ministry of Education, Culture, Sports, Science, and Technology.

%\bibliography{aps9.bib}% Produces the bibliography via BibTeX.

\end{document}